\documentclass[aps,10pt,onecolumn,amssymb,amsmath,nofootinbib,floatfix,notitlepage]{revtex4-1}
\usepackage{graphicx}
\usepackage{lmodern}
\usepackage[utf8]{inputenc}
\usepackage[UKenglish]{babel}
\usepackage{url}
\usepackage{hyperref}
\usepackage{amssymb}
\usepackage{xspace}
\usepackage[capitalise]{cleveref} 

\newcommand{\jbca}{{Jodrell Bank Centre for Astrophysics, School of Physics and Astronomy, The University of Manchester, Manchester M13 9PL, UK}}

\begin{document}
\author{Ian Harrison}
\email{ian.harrison-2@manchester.ac.uk}
\affiliation{\jbca}
\author{Michael L. Brown}
\email{m.l.brown@manchester.ac.uk}
\affiliation{\jbca}
\title{SKA Engineering Change Proposal:\\Gridded Visibilities to Enable Precision Cosmology with Radio Weak Lensing.}
\date{\today}
\begin{abstract}
This document was submitted as supporting material to an Engineering Change Proposal (ECP) for the Square Kilometre Array (SKA). This ECP requests gridded visibilities as an extra imaging data product from the SKA, in order to enable bespoke analysis techniques to measure source morphologies to the accuracy necessary for precision cosmology with radio weak lensing. We also discuss the properties of an SKA weak lensing data set and potential overlaps with other cosmology science goals.
\end{abstract}
\maketitle
\section{Radio Weak Lensing Cosmology}
Probing cosmology with the weak gravitational lensing effect is potentially the most sensitive way of determining the evolution of the Dark Energy equation of state $w$ with redshift and hence determining its physical nature. Weak lensing is therefore a primary science driver for high-profile upcoming optical experiments such as Euclid and LSST. However, weak lensing requires exquisitely precise measurements and a number of irreducible systematic effects may ultimately limit the power of these optical surveys.
\begin{figure}
\includegraphics[width=0.9\textwidth]{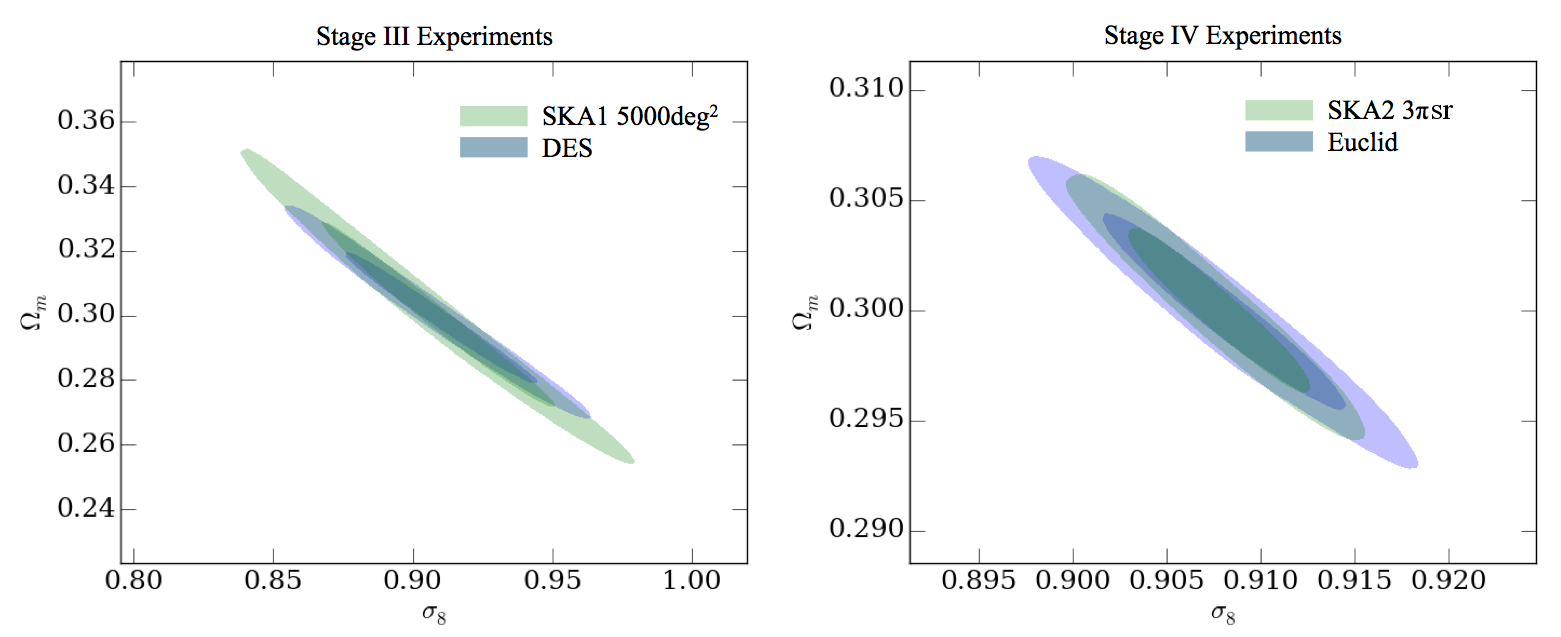}
\caption{Forecast constraints on total matter content $\Omega_{\mathrm{m}}$ and clustering $\sigma_{8}$ from stage III (left) and stage IV (right) weak lensing experiments, assuming source shapes can be measured to the same accuracy in different wavebands.}
\label{fig:forecasts}
\end{figure}
Large, deep surveys with SKA-MID will also be capable of achieving the number density of high redshift sources necessary for weak lensing, with both SKA1 and the full SKA (SKA2) providing constraints on cosmological parameters competitive with optical experiments, as shown in \cref{fig:forecasts}. In addition, there are a number of unique advantages to performing weak lensing in the radio (see \cite{2015arXiv150103828B} for more detail):
\begin{itemize}
\item Highly stable and deterministic point spread functions.
\item Source redshift distributions with higher median redshifts, giving more constraining power on evolution of Dark Energy and other cosmological parameters over cosmic time.
\item Extra information from polarisation and spectral line velocity measurements which may be used to remove otherwise limiting astrophysical systematics such as intrinsic galaxy alignments.
\item Cross correlations between optical and radio lensing maps may be expected to have uncorrelated telescope systematics, allowing the best unbiased constraints.
\end{itemize}
However, weak lensing has stringent and specific requirements on image
fidelity; the cosmological shear signal which affects source
ellipticity is small and so must be averaged over many source
shapes. This means source ellipticities must be measured with a systematic error of less than 1 part in 10,000 in order for errors on cosmological parameters to be dominated by statistical, rather than systematic, errors.

\begin{figure}[h!]
\includegraphics[width=0.4\textwidth]{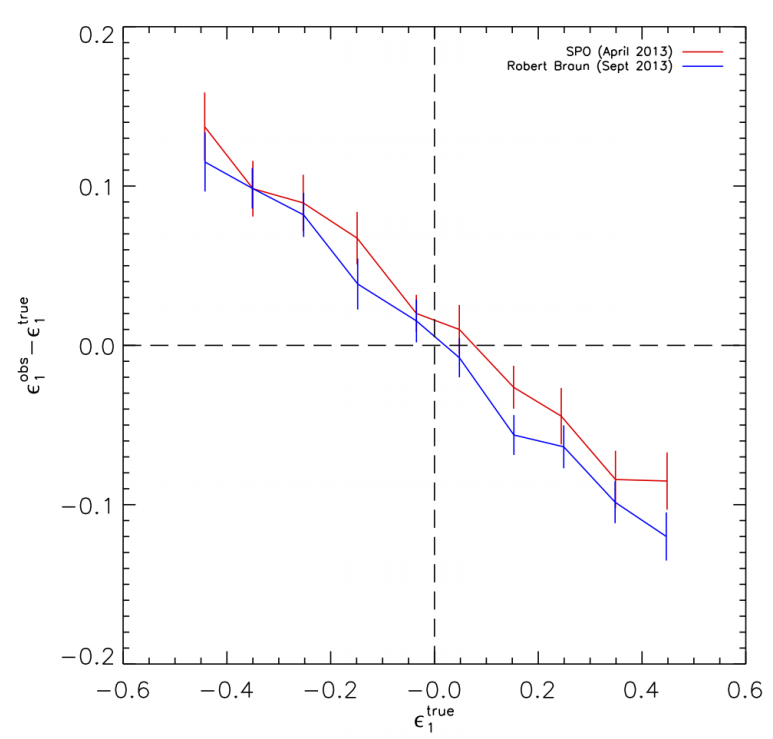}\\
\begin{tabular}{l|cc}
& Multiplicative Bias & Additive Bias\\
\hline
SKA1 requirement & $0.0067$ & $0.00082$\\
CLEAN images & $-0.265 \pm 0.02$ & $0.001 \pm 0.005$\\
\hline
\end{tabular}
\caption{Biases on ellipticity from CLEAN images of a simulated SKA1-MID data set from \cite{2015arXiv150103892P}. Top: input ellipticity plotted against difference to recovered ellipticity, showing the considerable multiplicative (slope) and additive (offset) biases. Bottom: requirements on biases for a $5000\,\mathrm{deg}^2$ SKA1 survey to have errors dominated by statistics, along with values recovered from CLEAN images.}
\label{fig:mandc}
\end{figure}

Simulations have shown that, even on high SNR objects (whose shapes
are easiest to measure) with an SKA1 antenna configuration, current
iterative deconvolution methods produce images with structures in the
residuals which dominate the cosmological signal, as shown in
\cref{fig:mandc} from \cite{2015arXiv150103892P}. The forecasted
constraints displayed in \cref{fig:forecasts} assume that shapes can
be measured with equivalent accuracy in the radio as is currently
possible in the optical. However, using ellipticity measurements with
the properties of those in \cref{fig:mandc} would produce constraints
with analysis-induced biases more than 100\% of the target measurement
of $w=-1$, far larger than the achievable 1\% statistical uncertainty.

We stress that this bias is \emph{not} a fundamental limitation of the SKA telescope, which will be capable of capturing all of the information necessary for precision weak lensing, but is induced by the analysis procedure adopted for turning the visibility data into images. This ECP requests that gridded visibilities be made available, which will facilitate analyses more optimal for the particular needs of weak lensing, thus enabling this frontier science.

The current SKA Phase 1 Level 1 requirements document provides for: moment images of a multi-frequency synthesis imaging process as the data products from the continuum imaging pipeline; and spectral line and continuum model images from the spectral line emission pipeline. In this ECP we request as an additionaly available data product the gridded visibilities which constitute the Fourier-plane representation of the dirty map, prior to the final imaging step. This will allow use of both bespoke imaging techniques for weak lensing, which may be tailored to the particular needs of this science case, and the potential for directly modelling the visibility plane data (as was done in the detection of a radio weak lensing signal by \cite{2004ApJ...617..794C}).

\section{Requirements for SKA}
Whilst gridded visibilities do not represent the full amount of information available from the raw visibilities, the operations performed when gridding should be either directly computed as part of the analysis pipeline (e.g. sampling, weighting, convolution with gridding kernel) or precisely known from features of the telescope (convolution with A and W-projection kernels), meaning their effect on source ellipticities should be able to be modelled. This is in contrast to the effect of an iterative deconvolution imaging algorithm on an unknown sky.

Requirements on these steps which move from raw to gridded visibilities will take the form of adequate samplings of the spatial scales of interest for weak lensing source modelling; adequate removal of bright confusing sources in far sidelobes; adequate w-projection and adequate primary beam calibration; and ensuring relevent morphological information is not destroyed by time and bandwidth smearing (as discussed below). Whilst it may be the case that many of these requirements will be satisfied along with other science cases requiring high dynamic range imaging, we expect to carry out detailed simulations to confirm exact requirements on the levels of sampling required. However, even within current default specifications for the gridding process, the ability to perform tailored imaging and source modelling analyses from gridded visbilities will be essential for enabling precision cosmology with radio weak lensing.

\begin{figure}
\includegraphics[width=0.45\textwidth]{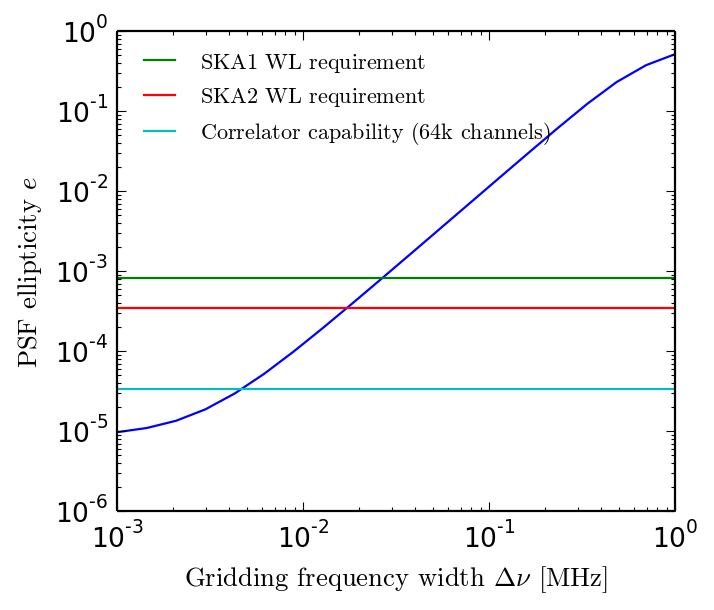}\includegraphics[width=0.45\textwidth]{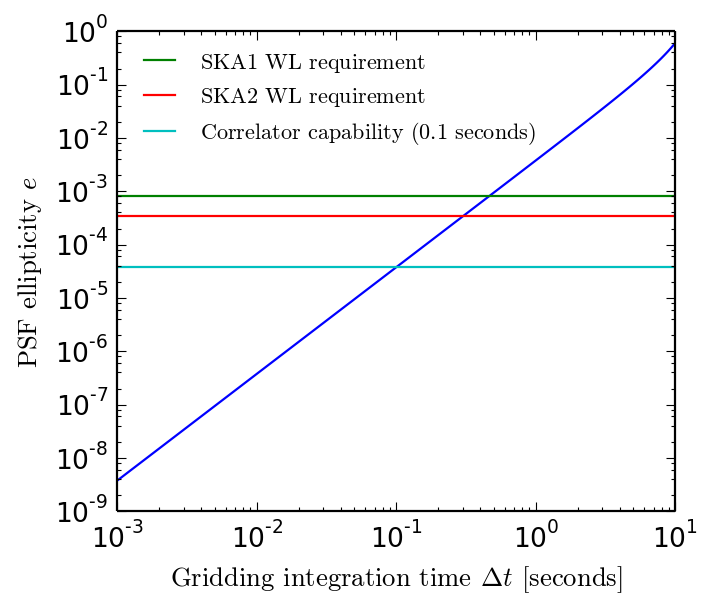}
\caption{Requirements on bandwidth (left) and time (right) smearing for systematic errors on measured ellipticity to be below statistical errors for SKA1 and SKA2 weak lensing surveys at the edige of the primary field of view ($25$ arcmin from the pointing centre).}
\label{fig:ellip_reqs}
\end{figure}

In order to get a sense of the necessary data set size for a weak lensing survey, we calculate the effect of time and bandwidth smearing on the ellipticities of sources across the field of view. We follow \cite{1999ASPC..180..371B} and calculate the change in the FWHM of the sythesised beam in the radial (tangential) direction for bandwidth (time) smearing at the edge of the field of view ($25$ arcmin from the pointing centre) for an SKA1-MID experiment with a circular synthesised beam FWHM at the pointing centre of $0.5\,\mathrm{arcsec}$. We vary the channel width and compare the maximum induced ellipticity (at the edge of the field of view) to requirements on the systematic uncertainty (as described in \cite{2015arXiv150103828B, 2015arXiv150103828B}) on PSF ellipticity for SKA1 and SKA2. The result is shown in the left panel of \cref{fig:ellip_reqs}, finding $\sim 50\,\mathrm{kHz}$ channels are required for the bandwidth smearing to be tolerable; for an observation making use of 30\% of the total available bandwidth (i.e. $300\,\mathrm{MHz}$), this means $\sim 6000$ channels will be required. Similarly for the time smearing, the right panel of \cref{fig:ellip_reqs} shows integration times $\sim 1$ second will be required for smearing-induced ellipticity to be at acceptable levels.

Whilst the smearing from both of these effects should be deterministic
and hence correctable, assessing the residuals from such corrections
across the full space of observations would require full-scale
(i.e. on the same scale as the data) visibility plane simulations. In
this case, the requirements on smearing would then be in terms of
required residuals after correction (which will likely be limited by
the variations in smearing across the size of a typical
source). However, generating and processing the massive simulations
which such a correction scheme would require represents a huge
computational challenge, equivalent to the collection and processing of
the real survey data. It therefore makes much more sense to place
requirements on the averaging procedures applied during the data processing, thus
ensuring that these systematic effects are suppressed to acceptable
levels in the initial data products without further corrections.

\section{Summary}
We expect this request to necessarily require no new algorithm development on the part of the SKA Organisation alone, with this being the responsibility of the science case. The impact on the telescope operations are expected to be as follows:
\begin{itemize}
  \item An addendum to the spectral line emission or continuum data pipeline which outputs as a data product the full gridded visibility cube (before FFT to a dirty image cube and deconvolution)
  \item Requirement for a larger data archive for observations comprising the  weak lensing survey. For a $5000\,\mathrm{deg}^{2}$ survey to a depth of $2\,\mu\mathrm{Jy}$ RMS, we expect to use $\sim10^4$ pointings of $\sim1$ hours each, requiring $\sim6000$ frequency channels at a resolution of $50\,\mathrm{kHz}$. This is a large multiplicative facor in data volume compared to a continuum survey, but is directly comparable to that expected by HI line galaxy surveys, who specify $10\,\mathrm{kHz}$ channels over similar total bandwidth and total observing time \cite{2015MNRAS.450.2251Y}.
  \item \emph{Alternatively}, a weak lensing dedicated pipeline, or sub-pipeline, which is capable of measuring source shapes in cached visibilties to the required level of accuracy.
  \item A corresponding increase in data transport from the Science Data Processor to the science analysis team.
  \item An increase in complexity for the Telescope Management system to allow for this mode of data output.
\end{itemize}
We expect this weak lensing survey to have similar requirements in terms of time, depth and footprint to other large, deep surveys at L-band (such as SKA1 Science Goal 27: ``The resolved all-Sky characterisation of the interstellar and intergalactic magnetic fields'', SKA1 Science Goal 33: ``Angular correlation functions to probe non-Gaussianity and the matter dipole'' and SKA1 Science Goal 36: ``Test dark energy \& general relativity with fore-runner of the `billion galaxy' survey.'' \cite{spo}), other users of which will benefit from the availability of gridded visibilities as a data product.

The SKA telescope will have the capacity to revolutionise the field of
weak lensing by extending the reach of this key cosmology experiment
to the radio band. Not only will the SKA be competitive with
state-of-the-art optical experiments (such as Euclid and LSST) in
terms of raw precision, it will offer truly unique approaches to weak
lensing (that are not possible with Euclid or LSST) through
polarisation and HI rotational velocity measurements. In order to
realise the huge science return available, bespoke galaxy shape
analyses will need to be applied to the data. The gridded visibilities
and/or weak lensing optimised sub-pipeline that we request in this
ECP will facilitate the application of such bespoke analyses, thus
ensuring that precision weak lensing cosmology remains within the
grasp of the SKA.  

\begin{acknowledgments}
MLB is an STFC Advanced/Halliday fellow. MLB and IH are supported by an ERC Starting Grant (grant no. 280127). The authors wish to thank Rosie Bolton, Keith Grainge, Anna Scaife, Tim Cornwell and Jeff Wagg for illuminating discussions and Lance Miller, Prina Patel, Marzia Rivi, Ian Heywood and Tom Kitching for helpful comments on the draft. This work made use of the CosmoSIS cosmological parameter estimation code, available at this URL: \url{https://bitbucket.org/joezuntz/cosmosis/wiki/Home} \cite{2015A&C....12...45Z}.
\end{acknowledgments}

\bibliography{supporting_doc}{}

\begin{thebibliography}{7}
\providecommand{\natexlab}[1]{#1}
\providecommand{\url}[1]{\texttt{#1}}
\providecommand{\urlprefix}{URL }
\providecommand{\eprint}[1][]{\url{#1}}

\bibitem[{{Braun} et~al.(2014){Braun}, {Bourke}, {Green} \& {Wagg}}]{spo}
{Braun}, R., {Bourke}, T., {Green}, J., {Wagg}, J., 2014, {SKA1 Science
  Priority Outcomes}, Tech. Rep. SKA-TEL-SKO-0000122, SKAO

\bibitem[{{Bridle} \& {Schwab}(1999)}]{1999ASPC..180..371B}
{Bridle}, A.~H., {Schwab}, F.~R., 1999, in Synthesis Imaging in Radio Astronomy
  II, edited by {Taylor}, G.~B., {Carilli}, C.~L., {Perley}, R.~A., vol. 180 of
  \emph{Astronomical Society of the Pacific Conference Series}, 371

\bibitem[{{Brown} et~al.(2015){Brown}, {Bacon}, {Camera}
  et~al.}]{2015arXiv150103828B}
{Brown}, M.~L., {Bacon}, D.~J., {Camera}, S., et~al., 2015, ArXiv e-prints,
  \eprint arXiv:{1501.03828}

\bibitem[{{Chang} et~al.(2004){Chang}, {Refregier} \&
  {Helfand}}]{2004ApJ...617..794C}
{Chang}, T.-C., {Refregier}, A., {Helfand}, D.~J., 2004, \apj, 617, 794,
  \eprint{astro-ph/0408548}

\bibitem[{{Patel} et~al.(2015){Patel}, {Harrison}, {Makhathini}
  et~al.}]{2015arXiv150103892P}
{Patel}, P., {Harrison}, I., {Makhathini}, S., et~al., 2015, ArXiv e-prints,
  \eprint arXiv:{1501.03892}

\bibitem[{{Yahya} et~al.(2015){Yahya}, {Bull}, {Santos}
  et~al.}]{2015MNRAS.450.2251Y}
{Yahya}, S., {Bull}, P., {Santos}, M.~G., et~al., 2015, MNRAS, 450, 2251,
  \eprint arXiv:{1412.4700}

\bibitem[{{Zuntz} et~al.(2015){Zuntz}, {Paterno}, {Jennings}
  et~al.}]{2015A&C....12...45Z}
{Zuntz}, J., {Paterno}, M., {Jennings}, E., et~al., 2015, Astronomy and
  Computing, 12, 45, \eprint arXiv:{1409.3409}

\end{thebibliography}
\bibliographystyle{mn2e_plus_arxiv}
\end{document}